\documentclass[conference]{IEEEtran}
\IEEEoverridecommandlockouts
\usepackage{cite}
\usepackage{amsmath,amssymb,amsfonts}
\usepackage{bm}
\usepackage{algorithmic}
\usepackage{graphicx}
\usepackage{textcomp}
\usepackage{tikz}
\usetikzlibrary{spy}
\usepackage{pgfplots}
\usepackage{xcolor}

\pgfplotsset{grid style={dotted,black}}
\pgfplotsset{compat=newest}
\def\BibTeX{{\rm B\kern-.05em{\sc i\kern-.025em b}\kern-.08em
    T\kern-.1667em\lower.7ex\hbox{E}\kern-.125emX}}
\begin{document}

\title{Learning from the Syndrome\\
\thanks{$^1$The first author performed this work while at McGill University.}
}

\author{\IEEEauthorblockN{Loren Lugosch$^1$}
\IEEEauthorblockA{
Fluent.ai Inc.\\
Montr\'{e}al, Qu\'{e}bec, Canada \\
loren.lugosch@fluent.ai}
\and
\IEEEauthorblockN{Warren J. Gross}
\IEEEauthorblockA{
McGill University\\
Montr\'{e}al, Qu\'{e}bec, Canada \\
warren.gross@mcgill.ca}
}
\maketitle

\begin{abstract}
In this paper, we introduce the syndrome loss, an alternative loss function for neural error-correcting decoders based on a relaxation of the syndrome. The syndrome loss penalizes the decoder for producing outputs that do not correspond to valid codewords. We show that training with the syndrome loss yields decoders with consistently lower frame error rate for a number of short block codes, at little additional cost during training and no additional cost during inference. The proposed method does not depend on knowledge of the transmitted codeword, making it a promising tool for online adaptation to changing channel conditions.
\end{abstract}


\section{Introduction}
Researchers are currently exploring the use of neural networks in digital communication systems, either as replacements for certain components or as an end-to-end solution. Much of this work has focused on trying to train improved decoders for error-correcting codes, since for many codes and channels the design of a near-optimal decoder is an unsolved problem. 
There have been many attempts to build machine learning-based decoders over the years \cite{an_application, 116658, SDD_ANN, 485019, 765550, abdelbaki1999random, wu2002neural}, but these attempts have largely been thwarted by the ``curse of dimensionality'' described in \cite{tenbrink}: for a code with $k$-bit messages, there are $2^k$ possible codewords, and a na\"ively configured learning algorithm may not generalize to the many codewords not seen during training. Indeed, \cite{176685} found that a fully-connected neural network for decoding even the simple (7,4) Hamming code could not successfully decode received vectors corresponding to codewords never shown during training. 

A few recent breakthroughs have reignited interest in the idea of decoding using deep learning. First, in \cite{LDDL}, Nachmani \textit{et al.} showed that by unrolling the belief propagation decoding algorithm for a number of iterations and assigning learnable weights to each iteration, a neural network is formed that can be trained to achieve error correction performance significantly better than that of the conventional belief propagation decoder for short high-density parity-check (HDPC) codes. Since the code is hardwired into the neural network structure, it suffices to train it using only the all-zeros codeword, thus sidestepping the curse of dimensionality. Subsequent works modified Nachmani \textit{et al.}'s approach to be more hardware-friendly \cite{NOMS}, use fewer parameters through weight sharing and attain close to optimal performance by being combined with a state-of-the-art HDPC decoder \cite{nachmani2017rnn, nachmani2018deep}, and handle channels with correlated noise using a convolutional neural network \cite{BP_CNN}. Second, in \cite{tenbrink}, Gruber \textit{et al.} reported the same failure of fully-connected neural networks to generalize to new codewords as was originally reported in \cite{176685}, but found that the effect was much less pronounced for codes for which the parity-check matrix is not random but rather has structure (specifically, polar codes \cite{arikan2009channel}), suggesting that fully-connected neural networks are capable of learning something like a decoding algorithm rather than simply memorizing the code. The approaches in \cite{doan2018neural} and \cite{cammerer2017scaling} strike a balance between fully-connected neural networks and conventional decoding algorithms to achieve lower latency decoding of polar codes.

Existing methods for training neural channel decoders have typically used the binary cross-entropy as the loss function for supervised learning. The cross-entropy loss is indeed an appropriate loss function for training a binary classifier. However, error correction is not a simple binary classification problem but rather a structured prediction problem, since the bits to be predicted are related to each other through the code structure. We therefore hypothesize that decoder training can be improved by incorporating knowledge of the code structure into the loss function. 

To test this hypothesis, we introduce a new loss function, the syndrome loss, which penalizes the decoder for producing outputs that do not correspond to valid codewords. We show that combining the syndrome loss with the cross-entropy loss improves the frame error rate of several neural channel decoders for short block codes across all signal-to-noise ratios.

Perhaps more interestingly, the syndrome loss is completely unsupervised: that is, the decoder does not require knowledge of the transmitted codeword in order to compute the loss. Unsupervised learning could enable online training of decoders without the use of pilot signals, a useful property for receivers that must adapt quickly to changing channel conditions \cite{schibisch2018online}.  We show that, while taking care not to overfit to the training codewords, decoders can indeed be trained using only unsupervised learning.

In the rest of the paper, we define the syndrome loss, relate it to previous work, and show how it may be useful using a set of supervised and unsupervised learning experiments.

\section{The Syndrome Loss}\label{syndrome_loss}
In this work, we consider communication systems that use a binary linear code to transmit over an additive white Gaussian noise (AWGN) channel, although our method could potentially be applied to other types of channel. The transmitter encodes a $k$-bit message $\bm{u} \in \text{GF}(2)^k$ using a generator matrix $\bm{G} \in \text{GF}(2)^{n \times k}$ to obtain an $n$-bit codeword $\bm{c} = \bm{Gu}  \in \text{GF}(2)^n$. The codeword is put in a bipolar format $\bm{x} = 1 - 2\bm{c} \in \{-1,+1\}^n$ and transmitted over the channel. The receiver receives a noisy signal $\bm{y} = \bm{x} + \bm{w} \in \mathbb{R}^n$, where $\bm{w} \in \mathbb{R}^n$ is a vector of AWGN channel noise with variance $\sigma^2$. The decoder must estimate $\bm{x}$ from $\bm{y}$. We consider decoders that produce a soft output $\bm{s} \in \mathbb{R}^n$, where the estimated bipolar codeword is found by taking the hard decision $\hat{\bm{x}} = \text{sign}(\bm{s})$, and the corresponding estimated binary codeword is $\hat{\bm{c}} = 0.5 - 0.5\hat{\bm{x}}$.

A linear code can be described by a parity-check matrix $\bm{H} \in \text{GF}(2)^{(n-k) \times n}$. For example, the following is a parity-check matrix for the (7,4)-Hamming code:
\begin{equation}\label{hamming}
	\bm{H} =     \begin{bmatrix}
        1       & 1       & 0         & 1        & 1 &    0 & 0         \\
        1       & 0       & 1         & 1         & 0 &    1 & 0    \\
        0       & 1       & 1         & 1         & 0 &    0 & 1 \\
    \end{bmatrix}.
\end{equation}

The product $\bm{H}\hat{\bm{c}} \in \text{GF}(2)^{n-k}$ is called the syndrome. If $\hat{\bm{c}}$ is a codeword, the syndrome will contain only $0$; otherwise, the syndrome will contain at least a single $1$. The syndrome can therefore be used to check if the decoder has successfully produced a valid codeword as an output.

Since adding numbers in $\text{GF}(2)$ is equivalent to multiplying numbers in $\{-1,+1\}$, the syndrome can be expressed equivalently in the bipolar format in terms of the soft output $\bm{s}$ as follows:
\begin{equation}
	\text{synd}(\bm{s})_i = \prod_{j \in \mathcal{M}(i)} \text{sign}(s_{j}),
\end{equation}
where $\mathcal{M}(i)$ is the set of columns in the $i$th row of $\bm{H}$ equal to $1$.

One could imagine training a decoder to produce outputs that are codewords by minimizing the number of entries of the syndrome equal to $-1$. However, the syndrome is not well suited for conventional gradient-based learning, since the gradient of each entry is $0$ almost everywhere. Accordingly, we introduce the ``soft syndrome'', a relaxation of the usual ``hard syndrome''. The soft syndrome is defined as follows:
\begin{equation}
	\text{softsynd}(\bm{s})_i = \min_{j \in \mathcal{M}(i)} |s_{j}| \prod_{j \in \mathcal{M}(i)} \text{sign}(s_{j}).
\end{equation}
Note that this is just the check node equation from min-sum decoding, which has a non-trivial gradient (c.f. Chapter 5 of \cite{lugosch2018learning}).

As an example illustrating the behavior of the soft syndrome, suppose that the transmitter using the (7,4)-Hamming code sends the all-zeros codeword,
\begin{equation}
	\bm{x} = \{+1,+1,+1,+1,+1,+1,+1\},
\end{equation}
and the receiver observes the sequence
\begin{equation}
	\bm{y} = \{+1.67, +1.42, \bm{-0.03}, +1.03, +0.88, +1.98, +0.44\},
\end{equation}
which contains one error. Suppose that the decoder outputs $\bm{s} = \bm{y}$. Whereas the hard syndrome given the parity-check matrix of Eq. \ref{hamming} evaluates to 
\begin{equation}
	\text{synd}(\bm{s}) = \{+1, \bm{-1} ,\bm{-1}\},
\end{equation}
the soft syndrome evaluates to
\begin{equation}
	\text{softsynd}(\bm{s}) = \{+0.88, \bm{-0.03} ,\bm{-0.03}\}.
\end{equation}

We can construct a loss function, the syndrome loss, based on the soft syndrome that penalizes all the entries that are negative as follows:
\begin{equation}
	\ell_{\text{syndrome}}(\bm{s}) = \frac{1}{n-k}\sum_{i=1}^{n-k} \max(1 - \text{softsynd}(\bm{s})_i,0).
\end{equation}

The usual supervised binary classification loss function 
is the cross-entropy loss:
\begin{equation}
	\ell_{\text{cross-entropy}}(\bm{c},\bm{s}) = \frac{1}{n}\sum_{j=1}^{n} c_j \log g(-s_j) + (1 - c_j) \log (1 - g(-s_j)),
\end{equation}
where $g(\cdot)$ is the logistic sigmoid function.

We propose to combine the syndrome loss with the cross-entropy loss to obtain a complete loss function:
\begin{equation}\label{eq:combined}
	\ell_{\text{total}}(\bm{c},\bm{s}) = (1-\lambda)\cdot \ell_{\text{syndrome}}(\bm{s}) + \lambda \cdot \ell_{\text{cross-entropy}}(\bm{c},\bm{s}),
\end{equation}
where $\lambda \in [0,1]$. When $\lambda = 1$, the loss is just the usual supervised loss; when $0 < \lambda < 1$, the loss is supervised with the syndrome loss as a regularization term; when $\lambda = 0$, there is no dependence on the transmitted codeword, so the loss is unsupervised.

\section{Related Work}
Other papers have proposed the use of something like the syndrome loss for decoding applications. In \cite{generalized_syndrome_weight}, the authors interpreted an iterative decoding algorithm as a gradient descent-based algorithm for minimizing a ``generalized syndrome weight''. This generalized syndrome weight was also treated in \cite{jiang2006iterative} and \cite{dimnik2009improved} in a similar way. In \cite{blind_LDPC_1} and \cite{blind_LDPC_2}, Xia and Wu approached the problem of blind detection and identification of LDPC codes using ``syndrome LLRs'' for each candidate code. 

It is important to distinguish our syndrome-based training method from that of \cite{bennatan2018deep}, in which the syndrome is calculated from the received signal and used as part of the input to a neural network decoder. In our method, the decoder can take on any form, as long as the output is a soft estimate of the transmitted codeword. Thus, our method is not suitable for decoders in which the output is an estimate of the original message $\bm{u}$, such as the polar decoder of \cite{tenbrink}. 

\section{Supervised Learning Experiments}
We trained neural normalized min-sum (NNMS) decoders \cite{nachmani2018deep} for four short block codes: a (63, 45) BCH code, a (16,8) LDPC code, a (128,64) polar code, and a (200,100) LDPC code. For all experiments described in this paper, we used the Adam update rule \cite{ADAM} with a learning rate of 0.01, and trained on 10,000 minibatches of 120 codewords each, with added noise drawn uniformly from all signal-to-noise ratios (SNRs). The all-zeros codeword was used during training, and random codewords were used during testing. We used the ``multi-loss'' approach proposed in \cite{LDDL}: the loss is computed for the soft output of every decoding iteration and these losses are summed to obtain the final loss.
We measured the frame error rate (FER) of the decoders using Monte Carlo simulation, requiring a minimum of 100 frame errors to be detected and at least 100,000 frames for each SNR to be simulated to minimize the variance of the FER estimates. The hyperparameter $\lambda$ was set to either 1 (purely supervised) or 0.5 (supervised $+$ regularized). Slightly better results can be obtained by tuning the value of $\lambda$; we have not attempted this here to show that the syndrome loss works even without careful tuning.

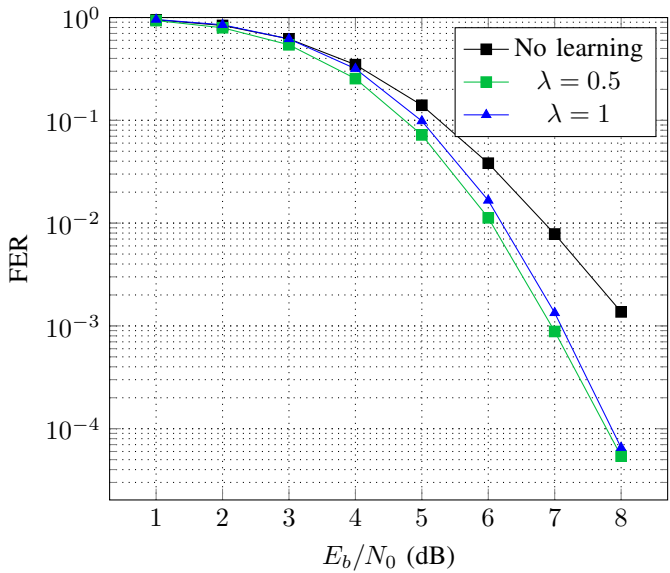
\begin{figure}
\centering
\begin{tikzpicture}[]
        \begin{semilogyaxis}[
            height=8cm,
            width=9cm,
            ymax=1,
            grid=both,
            xlabel=$E_b / N_0$ (dB),
            ylabel=FER
        ]
        
        \addplot[color=black,mark=square*] coordinates {
            (1,0.9503597122302159)
            (2, 0.8376898481215028)
            (3, 0.6196942446043165)
            (4, 0.34801159072741805)
            (5, 0.14015787370103916)
            (6, 0.038279376498800956)
            (7, 0.007833733013589129)
            (8, 0.0013689048760991208)
        };
        \addlegendentry{No learning}
        
        
        \addplot[color={rgb:red,0;green,3;blue,1},mark=square*] coordinates {
            (1, 0.9340327737809753)
            (2, 0.7955435651478817)
            (3, 0.5427757793764988)
            (4, 0.2545263788968825)
            (5, 0.0721123101518785)
            (6, 0.011250999200639489)
            (7, 0.0008827683615819209)
            (8, 5.41230975731203e-05)
        };
        \addlegendentry{$\lambda = 0.5$}
        
        \addplot[color={rgb:red,0;green,0;blue,3},mark=triangle*] coordinates {
            (1, 0.956724620304)
            (2, 0.849700239808)
            (3, 0.619314548361)
            (4, 0.317825739408)
            (5, 0.0984612310152)
            (6, 0.0166167066347)
            (7, 0.00133892885691)
            (8, 6.5298020164e-05)
        };
        \addlegendentry{$\lambda = 1$}

        \end{semilogyaxis}

    \end{tikzpicture}

\caption{Comparison of FER for decoders for the (63,45) BCH code trained with different values of $\lambda$.}
\label{fig:BCH_63_45_FER}
\end{figure}

\begin{figure}
\centering
\begin{tikzpicture}[spy using outlines=
	{circle, magnification=6, connect spies}]
        \begin{semilogyaxis}[
            height=8cm, 
            width=9cm,
            ymax=1,
            grid=both,
            xlabel=$E_b / N_0$ (dB),
            ylabel=FER
        ]
        
         \addplot[color=black,mark=square*] coordinates {
            (1, 0.9576438848920863)
            (2, 0.7004696243005596)
            (3, 0.23252398081534773)
            (4, 0.02180255795363709)
            (5, 0.0005807200929152149)
        };
        \addlegendentry{No learning}
        
        
        \addplot[color={rgb:red,0;green,3;blue,1},mark=square*] coordinates {
            (1, 0.92181254996)
            (2, 0.585411670663)
            (3, 0.145423661071)
            (4, 0.00927258193445)
            (5, 0.000189566272369)
        };
        \addlegendentry{$\lambda = 0.5$}
        
        \addplot[color={rgb:red,0;green,0;blue,3},mark=triangle*] coordinates {
            (1, 0.932434052758)
            (2, 0.611071143086)
            (3, 0.158882893685)
            (4, 0.010501598721)
            (5, 0.000203898540086)
        };
        \addlegendentry{$\lambda = 1$}
        
     \end{semilogyaxis}
    \end{tikzpicture}
    
\caption{Comparison of FER for decoders for the (200,100) LDPC code trained with different values of $\lambda$.}
\label{fig:LDPC_200_100_FER}
\end{figure}
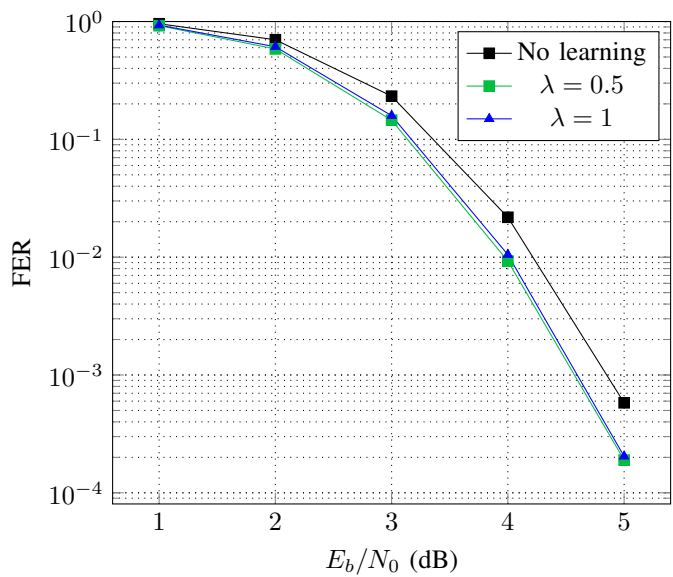

The performance of the decoders is shown in Fig. \ref{fig:BCH_63_45_FER}, \ref{fig:polar_128_64_FER}, \ref{fig:LDPC_16_8_FER}, and \ref{fig:LDPC_200_100_FER}. The performance for decoders without learning (i.e., all weights are equal to 1) is also shown for comparison. It can be seen that the decoders trained with the syndrome loss have a small but consistent improvement in FER across all signal-to-noise ratios. Thus, using the syndrome loss, decoders can be obtained with better FER performance at no additional cost during inference and little additional cost during training. The impact of the syndrome loss on bit error rate (BER), however, is less consistent. In some instances, we have found that BER is improved, and in others BER is made worse. It may be that the decoder attempts to output a valid codeword at the expense of making more bit errors.

\begin{figure}
\centering
\begin{tikzpicture}[spy using outlines=
	{circle, magnification=6, connect spies}]
        \begin{semilogyaxis}[
            height=8cm, 
            width=9cm,
            ymax=1,
            grid=both,
            xlabel=$E_b / N_0$ (dB),
            ylabel=FER
        ]
        
        \addplot[color=black,mark=square*] coordinates {
            (1, 0.9956235011990408)
            (2, 0.9678357314148681)
            (3, 0.861900479616307)
            (4, 0.6472621902478017)
            (5, 0.39813149480415666)
            (6, 0.20813349320543564)
            (7, 0.09914068745003997)
            (8, 0.03952837729816147)
        };
        \addlegendentry{No learning}
        
        
        \addplot[color={rgb:red,0;green,3;blue,1},mark=square*] coordinates {
            (1, 0.9921462829736211)
            (2, 0.9395883293365308)
            (3, 0.7429856115107913)
            (4, 0.3992406075139888)
            (5, 0.13218425259792166)
            (6, 0.029586330935251797)
            (7, 0.006684652278177458)
            (8, 0.0013689048760991208)
        };
        \addlegendentry{$\lambda = 0.5$}
        
        \addplot[color={rgb:red,0;green,0;blue,3},mark=triangle*] coordinates {
            (1, 0.9950639488409273)
            (2, 0.95646482813749)
            (3, 0.7859912070343725)
            (4, 0.4483812949640288)
            (5, 0.15331734612310152)
            (6, 0.03449240607513989)
            (7, 0.007394084732214229)
            (8, 0.001518784972022382)
        };
        \addlegendentry{$\lambda = 1$}
        
        \coordinate (spypoint) at (axis cs:7,0.001075);
        \coordinate (magnifyglass) at (axis cs:3,0.001);
        
        \end{semilogyaxis}
        

        
    \end{tikzpicture}

\caption{Comparison of FER for decoders for the (128,64) polar code trained with different values of $\lambda$.}
\label{fig:polar_128_64_FER}
\end{figure}
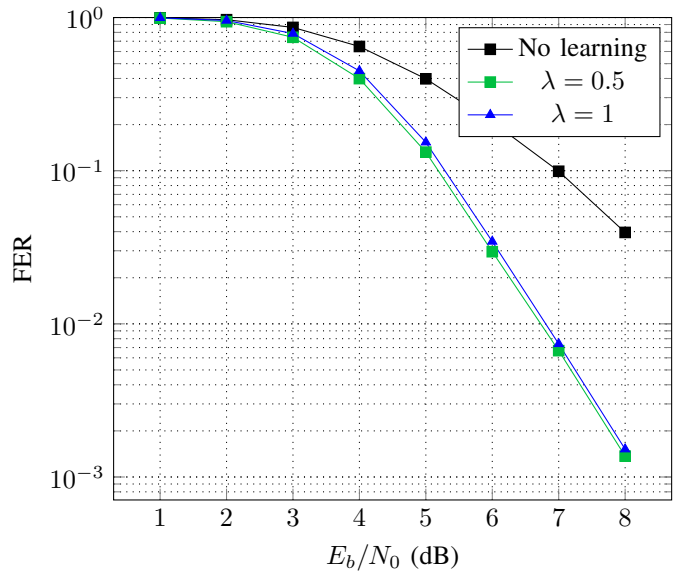

\begin{figure}
\centering
\begin{tikzpicture}[spy using outlines=
	{circle, magnification=6, connect spies}]
        \begin{semilogyaxis}[
            height=8cm, 
            width=9cm,
            ymax=1,
            grid=both,
            xlabel=$E_b / N_0$ (dB),
            ylabel=FER
        ]
        
        \addplot[color=black,mark=square*] coordinates {
            (1, 0.4634892086330935)
            (2, 0.31579736211031173)
            (3, 0.18246402877697843)
            (4, 0.08491207034372503)
            (5, 0.029596322941646682)
            (6, 0.00771382893685052)
            (7, 0.0011490807354116706)
            (8, 0.0001389352006224297)
        };
        \addlegendentry{No learning}
        
        \addplot[color={rgb:red,0;green,3;blue,1},mark=square*] coordinates {
            (1, 0.4230615507593925)
            (2, 0.27577937649880097)
            (3, 0.14947042366107113)
            (4, 0.06577737809752199)
            (5, 0.020123900879296563)
            (6, 0.004286570743405276)
            (7, 0.0005059704513256426)
            (8, 4.367804042839422e-05)
        };
        \addlegendentry{$\lambda = 0.5$}
        
        \addplot[color={rgb:red,0;green,0;blue,3},mark=triangle*] coordinates {
            (1, 0.46127098321342924)
            (2, 0.3108912869704237)
            (3, 0.17539968025579536)
            (4, 0.08001598721023181)
            (5, 0.025749400479616307)
            (6, 0.005955235811350919)
            (7, 0.0007716049382716049)
            (8, 6.283617352837681e-05)
        };
        \addlegendentry{$\lambda = 1$}
        
        \end{semilogyaxis}

    \end{tikzpicture}

\caption{Comparison of FER for decoders for the (16,8) LDPC code trained with different values of $\lambda$.}
\label{fig:LDPC_16_8_FER}
\end{figure}
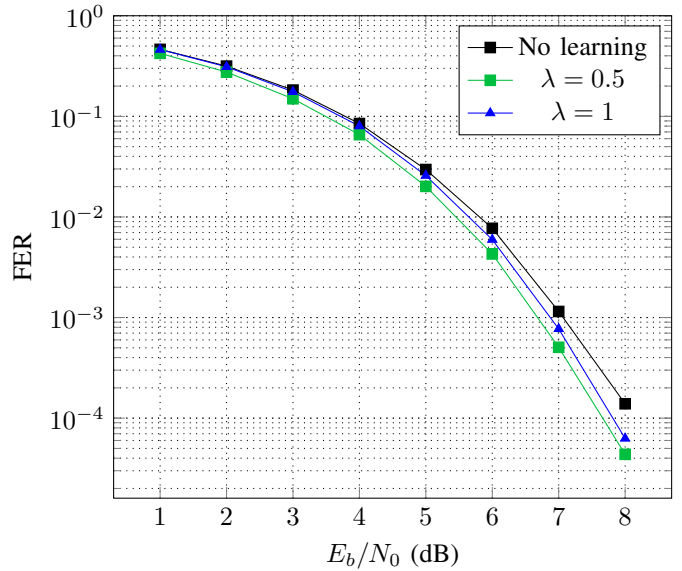

\section{Unsupervised Learning Experiments}
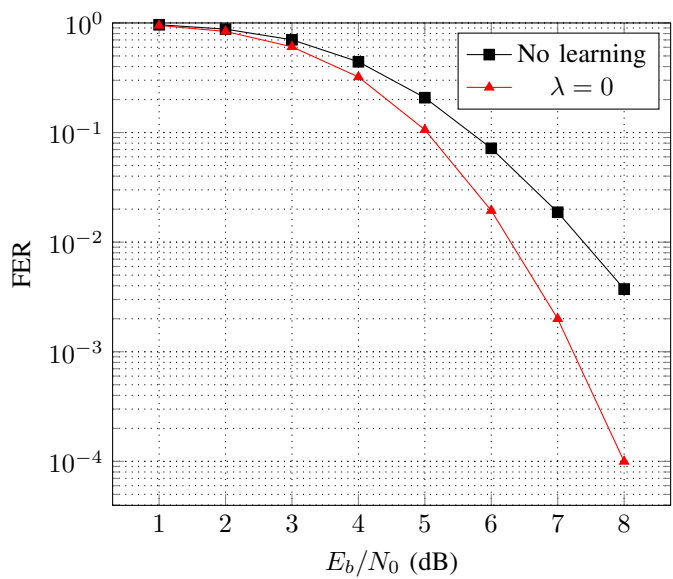
\begin{figure}
\centering
\begin{tikzpicture}[spy using outlines=
	{circle, magnification=6, connect spies}]
        \begin{semilogyaxis}[
            height=8cm, 
            width=9cm,
            ymax=1,
            grid=both,
            xlabel=$E_b / N_0$ (dB),
            ylabel=FER
        ]
        
        \addplot[color=black,mark=square*] coordinates {
            (1, 0.9625699440447641)
            (2, 0.8781374900079936)
            (3, 0.7020483613109513)
            (4, 0.4422961630695444)
            (5, 0.20803357314148682)
            (6, 0.07181254996003197)
            (7, 0.018745003996802557)
            (8, 0.003737010391686651)
        };
        \addlegendentry{No learning}
        
        \addplot[color={rgb:red,3;green,0;blue,0},mark=triangle*] coordinates {
            (1, 0.9486111111111111)
            (2, 0.836550759392486)
            (3, 0.6059452438049561)
            (4, 0.32147282174260594)
            (5, 0.1059652278177458)
            (6, 0.01940447641886491)
            (7, 0.0019984012789768186)
            (8, 9.971680427585657e-05)
        };
        \addlegendentry{$\lambda = 0$}
        
        \end{semilogyaxis}

    \end{tikzpicture}

\caption{Comparison of FER for decoders for the (63,36) BCH code  with or without unsupervised learning ($\lambda=0$).}
\label{fig:unsupervised}
\end{figure}

We attempted to train decoders using purely unsupervised learning, i.e. with $\lambda = 0$. In some instances, training the decoder with $\lambda = 0$ led to the decoder having $\text{FER} \approx 1$ across all SNRs. In these instances, because the decoder was trained using only the all-zeros codeword, it was able to find a set of positive and negative weights which, when multiplied with the all-zeros codeword, yield a valid (but incorrect) codeword. Two techniques were found to prevent this failure mode: 1) constraining the weights to being positive, e.g. using the softplus function (since we have observed that the weights are generally all positive after supervised learning), or 2) training using random codewords instead of the all-zeros codeword. The latter technique is preferable, since in theory some of the weights could be negative for the optimal parameter setting. The performance of an NNMS decoder for the (63, 36) BCH code trained on random codewords with $\lambda = 0$ is shown in Fig. \ref{fig:unsupervised}. The performance of the decoder with unsupervised learning is better than the decoder without learning, suggesting that the syndrome loss could potentially be used for online learning in decoders when the transmitted codewords are unknown.

While the syndrome loss does teach the decoder about the structure of the code, in principle there is no guarantee that this will help the decoder learn to decode. For example, a decoder that simply outputs a random codeword independent of the received signal would incur no syndrome loss. Therefore, some prior information about the goal of decoding must be provided to the decoder. For a neural belief propagation decoder, this information is built into the network through the graphical model of the code. For a \textit{tabula rasa} neural network, the prior information must be supplied in some other way, such as pre-training the model using supervised learning. We have not attempted to train an unconstrained neural network using the syndrome loss; we leave this for future work.


\section{Conclusion}

In this paper, we introduced the syndrome loss, a new loss function for neural error-correcting decoders. The syndrome loss is designed to teach decoders to produce outputs with a correct structure. Decoders trained using the syndrome loss have consistently lower frame error rate when the syndrome loss is used as a regularization term and are capable of purely unsupervised learning when the appropriate precautions are taken.

\section*{Acknowledgment}
Thanks to Ali Hashemi for providing the parity-check matrix for the polar code used in our experiments.

\bibliographystyle{IEEEtran}
\bibliography{references}


\end{document}